\documentstyle[12pt]{article}
\input epsf
\input epsf
\def\la{\mathrel{\mathpalette\fun <}}
\def\ga{\mathrel{\mathpalette\fun >}}
\def\fun#1#2{\lower3.6pt\vbox{\baselineskip0pt\lineskip.9pt
        \ialign{$\mathsurround=0pt#1\hfill##\hfil$\crcr#2\crcr\sim\crcr}}}

\def\sub#1{{{#1,_\phi}}}
\def\subsub#1{{{#1,_{\phi\phi}}}}
\def\subsubsub#1{{{#1,_{\phi\phi\phi}}}}

\def\be{\begin{equation}}
\def\ee{\end{equation}}
\def\bea{\begin{eqnarray}}
\def\eea{\end{eqnarray}}

\def\hij{{h_{ij}}}

\def\rr{{{\cal R}}}

\begin{document}
\begin{flushright}
\null\vspace{-72pt}
{\footnotesize
FERMILAB--Conf-96/437-A\\
astro-ph/9612138\\
December 1996 \\
Proceedings of Erice School\\
{\it Astrofundamental Physics}}
\end{flushright}

\vspace*{0.35in}

\begin{center}

{\bf INFLATION IN THE POSTMODERN ERA}

\vspace{0.5in}

EDWARD W.\ KOLB

{\it 
NASA/Fermilab Theoretical Astrophysics Group\\ 
Fermi National Accelerator Laboratory\\ 
Box 500, Batavia, Illinois 60510-0500, USA\\
and\\
Department of Astronomy and Astrophysics\\
Enrico Fermi Institute, The University of Chicago\\
5640 South Ellis Avenue, Chicago, Illinois 60637, USA}
\vspace{0.45in}

ABSTRACT
\end{center}
\begin{quote}
\small{ In this lecture I will review some recent progress in
improving the accuracy of the calculation of density perturbations
resulting from inflation.  }
\end{quote}

\section{Introduction}

The early universe was very nearly uniform.  However, the important
caveat in that statement is the word ``nearly.''  Our current
understanding of the origin of structure in the universe is that it
originated from small ``seed'' perturbations, which over time grew to
become all of the structure we observe.  The best guess for the origin
of these perturbations is quantum fluctuations during an inflationary
era in the early universe.

The basic idea of inflation is that there was an epoch early in the
history of the universe when potential, or vacuum, energy dominated
other forms of energy density such as matter or radiation. During the
vacuum-dominated era the scale factor grew exponentially (or nearly
exponentially) in time.  In this phase (known as the de Sitter phase),
a small, smooth spatial region of size less than the Hubble radius at
that time can grow so large as to easily encompass the comoving volume
of the entire presently observable universe.

If the early universe underwent this period of rapid expansion, then
one can understand why the universe is approximately smooth on the
largest scales, but has structure (people, planets, stars, galaxies,
clusters of galaxies, superclusters, etc.).  Inflation also predicts
that the cosmic background radiation should be very nearly isotropic,
with small variations in the temperature.  Perhaps all of the
structure we see in the universe is a result of quantum-mechanical
fluctuations during the inflationary epoch.  In this lecture I will
explore this possibility.

Because nearly all of the students are familiar with the basics of
cosmology, I will not bother to define familiar terms and notation.
In general, the notation follows that in {\it The Early Universe}
(Kolb and Turner, 1990), except that here the scale factor is denoted
by $a(t)$.

Since this is a school, I will not provide an exhaustive list of
references to original material, but refer to several basic papers
(including several review papers) where students can find the
references to the original material.  The list of references include
Bardeen (1980); Stewart (1990); Mukhanov, Feldman, and Brandenberger
(1992); Liddle and Lyth (1993); and Lidsey, Liddle, Kolb, Copeland,
Barreiro, and Abney (L$^2$KCBA) (1997).

\section{Evolution of Perturbations}

\subsection{Life Beyond the Hubble Radius}

An important part of this lecture will be the interplay of physical
length scales with the Hubble radius.  The time-dependent Hubble
radius is defined as the inverse of the expansion rate: $R_H(t) \equiv
H^{-1}(t) = [8\pi G \rho(t) /3]^{-1/2}$ (the last part of the equation
comes from the Friedmann equation for a spatially flat universe).  In
a radiation-dominated (RD) universe $\rho\propto a^{-4}$ and in a
matter-dominated (MD) universe $\rho\propto a^{-3}$, so $R_H\propto
a^{2}$ in an RD universe and $R_H\propto a^{3/2}$ in a MD universe.

First, let us review what is meant by ``crossing'' the Hubble radius.
For the sake of illustration, let's take a length scale $\lambda$ to
be at present $\lambda_0=300 h^{-1}$Mpc.  Today the Hubble radius is
$R_H(t_0)=H_0^{-1}\sim 3000 h^{-1}$Mpc, so
$\lambda_0/R_H(t_0)=10^{-1}$ and $\lambda_0$ is said to be ``within''
the Hubble radius today.  Any physical length scale increases in
proportion to the scale factor in an expanding universe. The scale
which is $\lambda_0$ today, was smaller in the early universe by a
factor of $a(t)/a_0=1/(1+z)$, where $a_0$ is the present scale factor.
The Hubble radius also depends upon $a(t)$, e.g.,
$R_H=[a(t)/a_0]^{3/2}=1/(1+z)^{3/2}$ in the MD era.  So during the MD
era the ratio $\lambda(t)/R_H(t)$ depends upon redshift as
$\lambda(t)/R_H(t) = [\lambda_0/R_H(t_0)] (1+z)^{1/2} =
10^{-1}(1+z)^{1/2}$.  So for $z\ga100$, the length scale $\lambda$ was
{\it outside} the Hubble radius, for $z\la100$, the length scale
$\lambda$ was {\it inside} the Hubble radius.  At $z=100$ we say that
a length scale of $300 h^{-1}$Mpc {\it crossed} the Hubble radius.

\begin{figure}
\centerline{ \epsfysize=235pt  \epsfbox{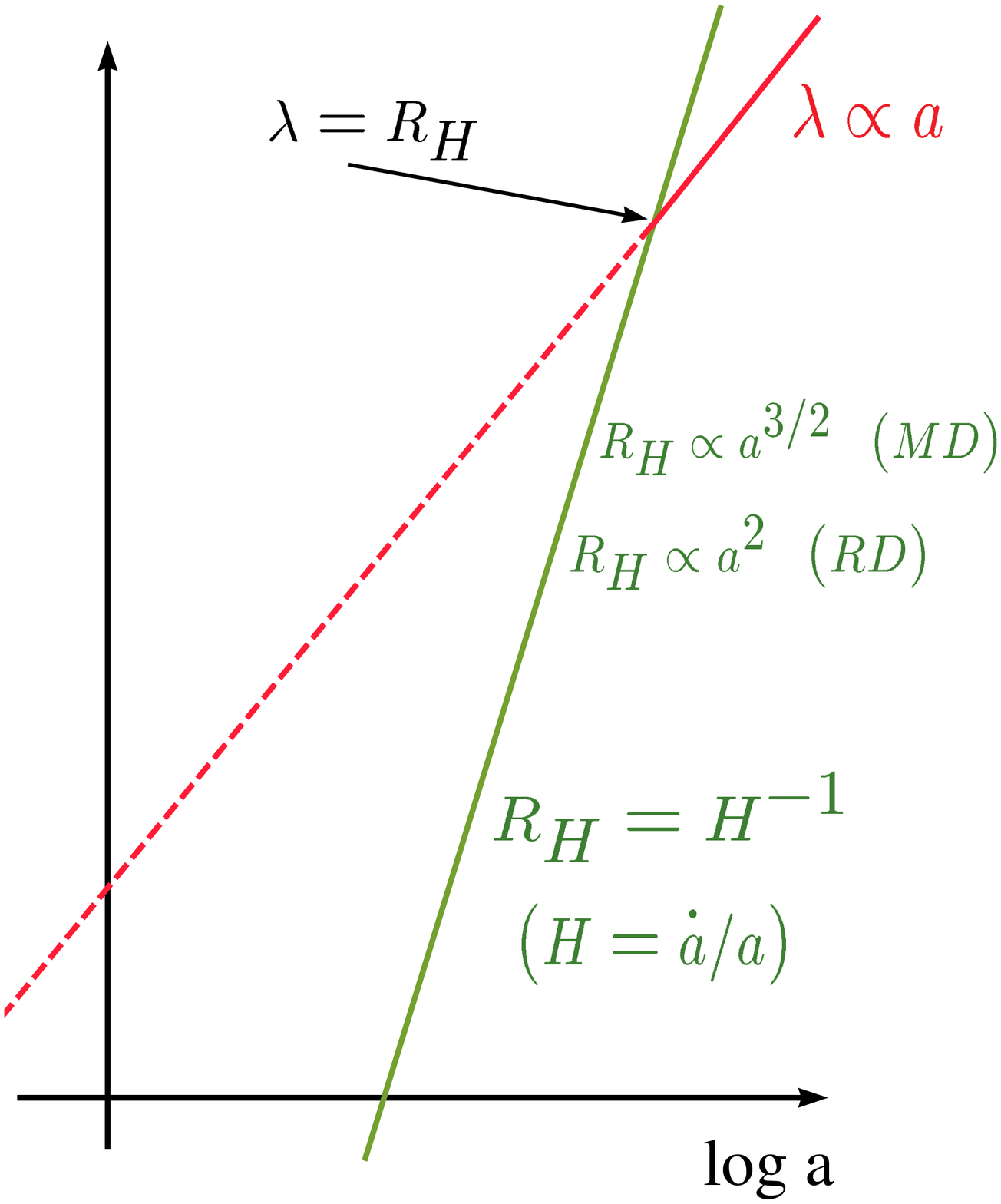} \epsfysize=235pt \epsfbox{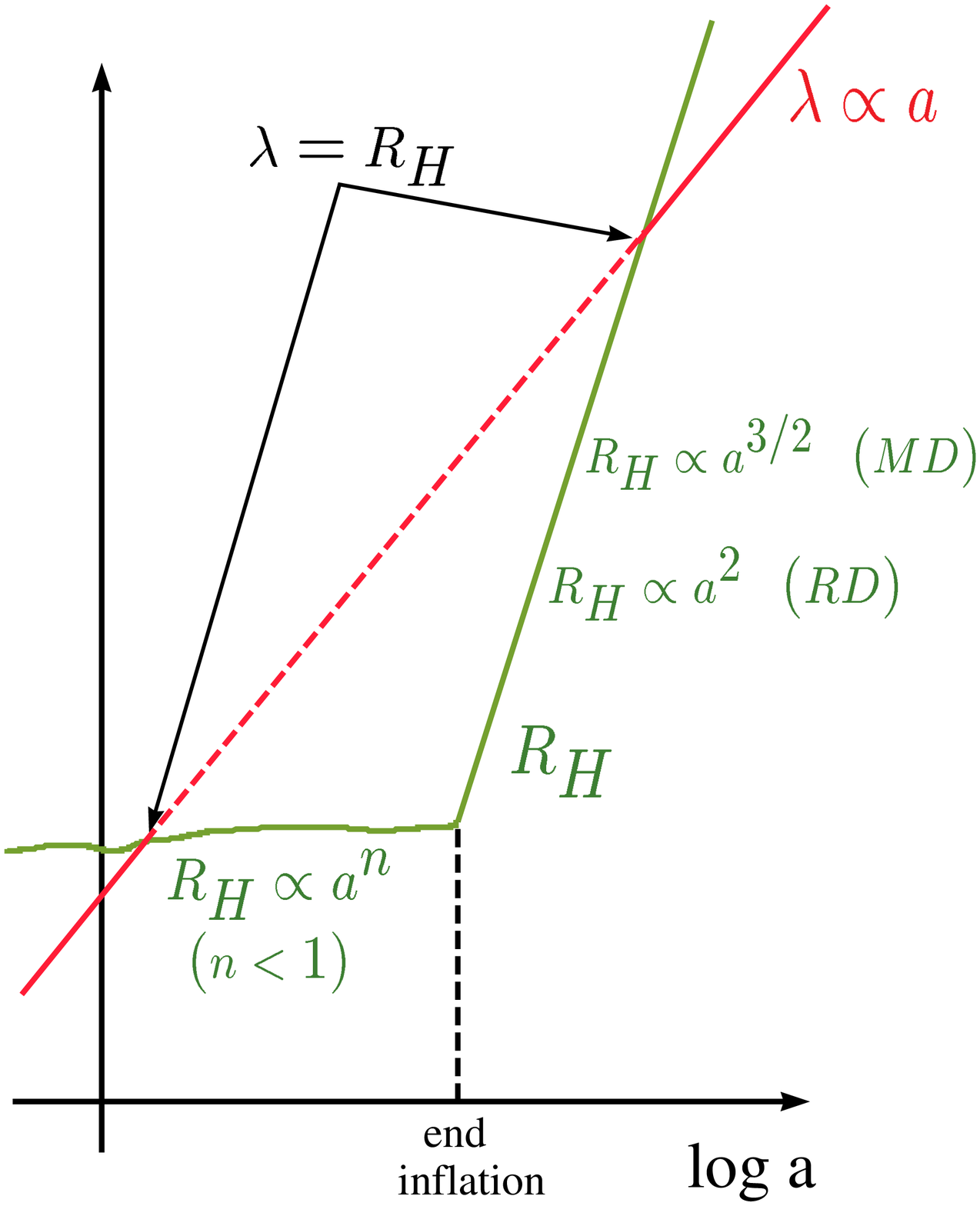} }
\caption{{ The
 behavior of the Hubble radius, $R_H$, and the physical size,
 $\lambda$, with the scale factor $a$ during normal expansion (left)
 and inflationary expansion (right).}}
\label{singledouble}
\end{figure}

Since $\lambda(t)/R_H(t) $ decreases in time in a radiation-dominated
or matter-dominated universe, any physical length scale $\lambda$
starts larger than $R_H$, then crosses the Hubble radius
($\lambda=H^{-1}$) only once.  This behavior is illustrated by the
left side of Fig.\ \ref{singledouble}.

In order for structure formation to occur via gravitational
instability, there must have been small preexisting fluctuations on
physical length scales when they crossed the Hubble radius in the RD
an MD eras.  In the standard big-bang model these small perturbations
have to be put in by hand, because it is impossible to produce
fluctuations on any length scale while it is larger than $R_H$.  Since
the goal of cosmology is to understand the universe on the basis of
physical laws, this appeal to initial conditions is unsatisfactory.

That any length scale crosses $R_H$ only once is not a fundamental
result of anything sacred like Einstein's equations, the cosmological
principle, or special relativity, but it depends upon the assumption
of the equation of state.  To see how changing the equation of state
changes the ratio $\lambda(t)/R_H(t)$, let's define $L(t)$ to be the
dimensionless ratio $\lambda(t)/R_H(t)$.  Obviously, if $L(t)$ is
smaller than unity, the scale is within the Hubble radius and it is
possible to imagine some microphysical process establishing
perturbations on that scale, while if $L(t)$ is larger than unity, it
is beyond the Hubble radius and no microphysical process can account
for perturbations on that scale.  Now $R_H(t) = H^{-1}(t)=
a(t)/\dot{a}(t)$ and $\lambda(t) \propto a(t)$, so $L(t)$ is
proportional to $\dot{a}(t)$, and $\dot{L}(t)$ scales as
$\ddot{a}(t)$, which from the Einstein equation is proportional to
$-(\rho+3p)$.  There are two possible scenarios for $\dot{L}(t)$
depending upon the sign of $\rho+3p$: 
\be \dot{L}(t) \left\{
\begin{array}{lll} 	< 0 \rightarrow &  R_H(t) 
\mbox{ grows faster than } \lambda(t), 
	& \mbox{occurs for }\rho+3p>0 \\
				> 0 \rightarrow &  R_H(t) 
\mbox{ grows more slowly than } \lambda(t), 
	& \mbox{occurs for }\rho+3p  <0.
\end{array}
\right. 
\ee

If during some epoch the equation of state was such that $\rho+3p <
0$, then scales larger than $R_H$ remained larger than $R_H$, while
scales smaller than the Hubble radius were destined eventually to grow
larger than the Hubble radius.  The opposite behavior obtains during
the standard RD and MD epochs when $\rho+3p > 0$.  During these epochs
scales smaller than $R_H$ remain smaller than $R_H$ and scales larger
than $R_H$ eventually become smaller than $R_H$.

Now if $\rho+3p < 0$ in the early universe and $\rho+3p > 0$ in the
later universe, then it is possible to have a ``double-cross''
situation illustrated on the right side of Fig.\ \ref{singledouble}.
In the double-cross scenario, length scales start smaller than the
Hubble radius during the phase when $\rho+3p < 0$ (the inflationary
phase), cross the Hubble radius, then remain larger than the Hubble
radius.  During the standard phase, scales of astrophysical interest
start larger than the Hubble radius, cross the Hubble radius, then
remain smaller than the Hubble radius.

Unlike the standard model, the double-cross model has the feature that
it is possible to imprint perturbations on a scale as it crosses the
Hubble radius during the inflationary phase, so one can imagine a
reason to have preexisting perturbations on scales recrossing the
Hubble radius during the RD-MD epochs.

\subsection{Metric Perturbations on Scales Larger than $R_H$}

\begin{figure}
\centerline{\epsfysize=300pt \epsfbox{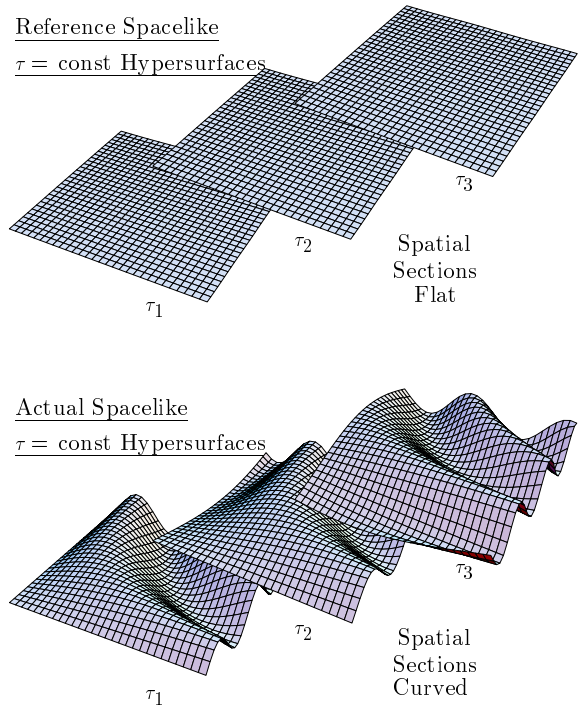}  }
\caption{{In the reference unperturbed universe, constant-time
surfaces have constant spatial curvature (zero for a flat FRW model).
In the actual perturbed universe, constant-time surfaces have
spatially varying spatial curvature.}}
\label{tau}
\end{figure}

What we are interested in following the evolution of a spacetime which
is neither homogeneous nor isotropic.  We will do this by following
the evolution of the differences between the actual spacetime and a
well understood reference spacetime.  So we will consider small
perturbations away from the homogeneous, isotropic spacetime (see
Fig.\ \ref{tau}).

When one studies ``perturbations'' it is necessary to specify a
reference background system.  The reference system in our case is the
spatially flat Friedmann--Robertson--Walker (FRW) spacetime, with line
element $ds^2 = a^2(\tau) \left\{ d\tau^2 - \delta_{ij} dx^idx^j
\right\}$, where $\tau$ is conformal time, related to ``normal'' time
by $a^2d\tau^2=dt^2$.  Sometimes equations will be written in terms of
conformal time $\tau$, and sometimes in terms of coordinate time $t$.
Derivatives with respect to conformal time will be denoted by a prime,
while usual time derivatives are denoted by a dot, e.g., the Hubble
parameter can be defined as $H(t) \equiv \dot{a}/a$, or ${\cal
H}(\tau) = a'/a = Ha$.

The most general form of a metric describing small perturbations away
from the flat FRW metric contains scalar, vector, and tensor
perturbations [the covariant decomposition of $\delta g_{\mu\nu}$ is
given in Stewart (1990)].  For the moment, we will only be interested
in the scalar perturbations.  The perturbed line element including the
scalar perturbations can be written in terms of four scalar functions
$\{A,\ B,\ \psi,\ E \}$: \be
\label{eq:SCALARP}
ds^2 = a^2(\tau)\left\{ (1+2A)d\tau^2 - 2\partial_iB\, dx^i d\tau -
\left[ (1-2\psi)\delta_{ij} + 2 \partial_i\partial_j E \right] dx^i
dx^j \right\} .  
\ee

Now because of the residual gauge freedom, not all of the four scalar
perturbation functions $\{A,\ B,\ \psi,\ E \}$ are independent.  For
instance if one works in the synchronous gauge, all hypersurfaces have
the same time.  In this gauge $A=B=0$, and the line element is
$ds^2=a^2(\tau) \left\{ d\tau^2 - \left[ (1-2\psi)\delta_{ij} + 2
\partial_i\partial_j E \right] dx^i dx^j \right\}$.  In the
longitudinal gauge $B=E=0$, and the line element is $ds^2=a^2(\tau)
\left\{ (1+2A)d\tau^2 - \left[ (1-2\psi)\delta_{ij} \right] dx^i dx^j
\right\}$.

It is sometimes bewildering to read the literature because everyone
seem to have his/her favorite gauge.  But really smart people support
freedom of choice, and work with combinations of the {\it
gauge-invariant} scalar functions $\Psi$ and $\Phi$ first found by
Bardeen (1980):
\begin{eqnarray}
\Psi & = & \psi -{\cal H}(B-E') \cr
\Phi & = & A + a^{-1}[(B-E')a]\, '   .
\end{eqnarray} 
Note that in the longitudinal gauge $\Phi = A$ and $\Psi=\psi$.

\subsection{Perturbations in the Stress-Energy Tensor}

Inflation assumes that the universe was dominated by something with an
equation of state that satisfies the inequality $\rho+3p<0$.  Such a
component of the energy density is usually called ``vacuum energy.''
Since today we know that the vacuum energy must be very small
(compared to what is required for inflation),\footnote{And tasteful
people assume that today the vacuum energy vanishes, i.e.,
$\Lambda=0$.} any inflationary model has to have some dynamics for
changing the vacuum energy.  It is convenient to imagine that the
dynamics of the change in the equation of state during inflation is
described by the usual dynamics of a minimally coupled scalar field
evolving under the influence of a scalar potential.  This mysterious
scalar field, denoted by $\phi$, is known as the {\it inflaton}, and
its potential, $V(\phi)$, is known as the {\it inflaton potential.}

One assumes that the inflaton field is homogeneous in the reference
spacetime, $\phi({\bf x},\tau)=\phi_0(\tau)$, and satisfies the
equation of motion $\ddot{\phi}_0+3H \dot{\phi}_0+\sub{V}=0$.  (Since
``prime'' was used to denote $d/d\tau$, I will use $X,_\phi$ to denote
$dX/d\phi$.)  This field equation, together with the Friedmann
equation, can be solved to find the evolution of the background
spacetime and scalar field.  Alternatively, one can view the scalar
field itself as the dynamical variable of the system.  This allows the
Einstein--scalar-field equations to be written as a set of
first-order, non-linear differential equations (Grishchuk \& Sidorav
1988; Muslimov 1990; Salopek \& Bond 1990, 1991; Lidsey 1991a)
\begin{eqnarray}
\label{eq:hj}
[\sub{H }(\phi)]^2 - \frac{12\pi}{m_{Pl}^2}H^2(\phi) & = & - 
\frac{32\pi^2}{m_Pl^4}V(\phi) \cr
\dot{\phi}  & = & - \frac{m_{Pl}^2}{4\pi}\sub{H}.
\end{eqnarray}

In the actual perturbed spacetime there are small perturbations about
the background value: $\phi({\bf x},\tau)=\phi_0(\tau) +
\delta\phi({\bf x},\tau)$.  Of course we will be interested in the
evolution of $\delta \phi$.

Now just as the metric perturbations are gauge dependent, the scalar
field fluctuations are also.  But one can circumvent the usual
problems associated with gauge freedom by constructing a suitable
gauge-invariant scalar field fluctuation, $\widetilde{\delta\phi} =
\delta\phi + \phi_0'(B-E')$.

\subsection{Perturbation Spectra}

Of great  convenience is the particular gauge-invariant quantity
\be 
\label{eq:rr}
\rr = \Psi - \frac{{\cal H}}{\phi_0'}\widetilde{\delta\phi} = \psi -
\frac{H}{\dot {\phi}}\delta\phi.  
\ee 
Clearly $\rr$ defined by the
first equality of Eq.\ (\ref{eq:rr}) is gauge invariant because it is
constructed explicitly from gauge-invariant terms.  However even the
second form in Eq.\ (\ref{eq:rr}) is gauge invariant, as when one
performs a gauge transformation the non-gauge-invariant terms in
$\psi$ cancels the non-gauge-invariant terms in the $\delta\phi$ term.

\begin{figure}
\centerline{\epsfysize=220pt  \epsfbox{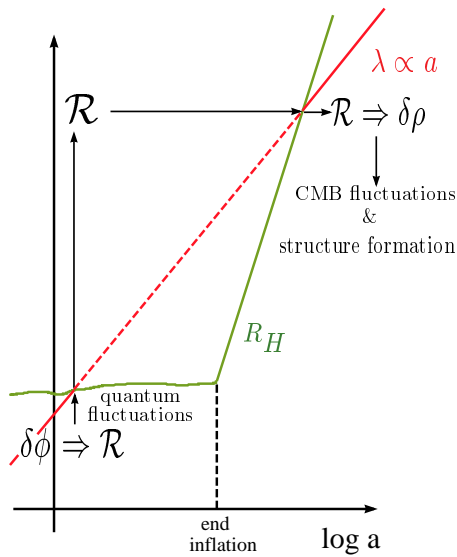} }
\caption{{Quantum fluctuations in the inflaton field during inflation
lead to fluctuations in the gauge-invariant quantity $\rr$.  As a
particular length scale passes outside of the Hubble radius, the
fluctuations are ``frozen in'' because $\rr$ is constant outside the
Hubble radius.  When the length scale reenters the Hubble radius in
the RD or MD era, fluctuations in $\rr$ appear as energy density
fluctuations. }}
\label{dphir}
\end{figure}

$\rr$ has a simple physical interpretation in the synchronous gauge,
where $\nabla^2\rr = {^3\!R}/4$ with $^3\!R$ the three-dimensional
Ricci curvature on the spatial hypersurface.  Now the usefulness of
$\rr$ follows from the fact that as shown by Bardeen (1980), $\rr$ is
constant on scales much larger than $R_H$.

The picture of the generation of quantum fluctuations during inflation
can be appreciated by studying Fig.\ \ref{dphir}.

Now $\rr$ is related to the observationally determined power spectrum.
The first step in developing the relation is to expand $\rr$ in terms
of Fourier modes $\rr_k$ 
\be \rr({\bf x}) =\int
\frac{d^3\!k}{(2\pi)^{3/2}} \rr_k(\tau) \ e^{i{\bf k} \cdot {\bf x}}.
\ee 
Now following the usual procedure, if we form $\langle {\cal
R}({\bf x}) {\cal R}({\bf x}) \rangle^{1/2}$, where $\langle \cdots
\rangle$ indicates the spatial average, we find that it is
proportional to $ \int k^2 |{\cal R}_k|^2 dk/k$, so $k^{3/2} |{\cal
R}_k| $ is the power in ${\cal R}$ per decade of $k$.  If the
curvature perturbation is independent of $k$, then the
``power-per-decade'' is constant, and $|{\cal R}_k| \propto k^{-3/2}$.
Putting in the factors of $2\pi$, we define the scalar spectrum
$A_S(k)$ by\footnote{The exact constant of proportionality is a matter
of convention, see L$^2$KCBA.}  
\be \frac{k^3}{2\pi^2} \ \langle \rr_k
\rr_l^* \rangle = \frac{25}{4}\ A_S^2(k) \ \delta^3({\bf k} - {\bf
l}), 
\ee 
where $A_S(k)$ is the {\it primordial} scalar density
perturbation power spectrum.  If ${\cal R}$ is independent of scale
outside of the Hubble radius, then $A_S(k)$ will be independent of
$k$.  The primordial power spectrum, $A_S(k)$, is related to $P_S(k)$,
the power spectrum {\it observed} in large-scale structure (LSS)
surveys and cosmic background radiation (CBR) experiments.

To find the relation between $A_S(k)$ and $P_S(k)$, it is important to
appreciate that $A_S(k)$ is the amplitude when a scale $k$ crosses the
Hubble radius, i.e., when $k=aH$.  Now if we specify the perturbation
spectrum on a particular space-like hypersurface, rather than as each
scale crosses the Hubble radius, we have to realize that we are
specifying a gauge-dependent quantity beyond the Hubble radius.  We
will denote the perturbation defined this way as
$(\delta\rho/\rho)_k$.  In the synchronous gauge and in the comoving
gauge, the density perturbation of wavenumber $k$ grows as
$(\delta\rho/\rho)_k\propto (aH)^{-2}$ for $k<aH$ in both the MD and
RD eras.  So for scales well outside the Hubble radius,
$(\delta\rho/\rho)_k \propto [k^2/(aH)^2]A_S(k)$, so that when $aH=k$,
$(\delta\rho/\rho)_k=A_S(k)$.

For scales inside the Hubble radius the synchronous gauge and the
comoving gauge coincide, and $(\delta\rho/\rho)_k$ is approximately
constant in the RD era and grows as $(\delta\rho/\rho)_k\propto
(aH)^{-2}$ in the MD era.  So just around the time of matter
domination, on scales smaller than $R_H$ (i.e., $k\gg (aH)_{EQ}$),
$(\delta\rho/\rho)_k$ has the approximate value it had when it crossed
the Hubble radius, so $(\delta\rho/\rho)_k \sim A_S(k)$ for $k\gg
(aH)_{EQ}$.  After matter domination $(\delta\rho/\rho)_k$ grows as
$(aH)^{-2}\sim a \propto t^{2/3}$ on all scales, so $
(\delta\rho/\rho)_k$ will continue to have the shape it did just after
matter domination (at least in the regime of linear evolution).

The transition between scales larger than $R_H$ at $t_{EQ}$ and scales
smaller than $R_H$ at $t_{EQ}$ can be encoded in a ``transfer
function'' $T(k)$, by writing $(\delta\rho/\rho)_k = (k/aH)^2T(k)
A_S(k)$ (see e.g., Liddle and Lyth, 1993).  In order to reproduce the
behavior discussed above, the function $T(k)$ must have the limiting
forms $T(k)\rightarrow1$ for $k\ll aH$ and $T(k)\rightarrow k^{-2}$
for $k\gg aH$.

Now the power spectrum $P_S(k)$ is defined by $(\delta\rho/\rho)_k^2
\propto k^3 P_S(k)$, so in terms of the primordial spectrum $A_S(k)$
and the transfer function $T(k)$, $P_S(k)$ is given by $P_S(k) \propto
k T^2(k) A_S^2(k)$.  Note that if the primordial spectrum is
independent of scale, i.e., if $A_S^2(k)$ is independent of $k$---the
Harrison-Zel'dovich spectrum, then $P_S(k) \propto k$ for $k\ll
(aH)_{EQ}$ and $P_S(k) \propto k^{-3}$ for $k\gg (aH)_{EQ}$.  If we
write $A_S(k)$ as a power law, $A_S^2(k) = A_S^2(k_0)(k/k_0)^{n-1}$,
then $P_S$ will be a power law also: $P_S(k) \propto k^n$, with $n=1$
corresponding to the value for constant amplitude perturbations at
Hubble radius crossing.

Finally, we must understand the relation between wavenumber $k$ and
field value $\phi$.  During the evolution of the scalar field the
background value of $\phi$ changes in time.  Now associated with a
particular value of $\phi$ is a length scale with comoving wavenumber
$k$ crossing the Hubble radius at the time the scalar field value is
$\phi$.  The easiest relation to find is the differential form found
from the expression $k=aH$: 
\be
\label{eq:kphi}
\frac{d\ln k}{d\phi} = \frac{\sub{H}}{H} + \frac{\sub{a}}{a} =  
\frac{\sub{H}}{H}  
- \frac{4\pi}{m_{Pl}^2}\frac{H}{\sub{H}},
\ee
where the last equality follows from Eq.\ (\ref{eq:hj}). 

\section{Scalar Perturbations from Inflation}

\subsection{Textbook Treatment}

In the standard textbook treatment one expands fluctuations of the
inflaton field in a Fourier expansion 
\be \langle
\delta\phi^2\rangle^{1/2} = \int \frac{dk}{k}\, k^3 \left| \delta_k
\right|^2.  
\ee 
Then one identifies $(\delta\phi)_k\equiv
k^{3/2}|\delta\phi_k|$ as the fluctuation in the inflaton on the
length scale $2\pi/k$.  One knows that a scalar field has quantum
fluctuations in deSitter space, or more precisely, the quantum
fluctuations of a scalar field in deSitter space differ from the
quantum fluctuations of a scalar field in flat space.  The quantum
fluctuations of a field in deSitter space at Hubble radius crossing
(i.e., on scales $k=aH$, where $k$ is the comoving wavenumber) result
in $(\delta\phi)_{k=aH} = H/2\pi$.

Now in the $\psi=0$ gauge, using the above result for the scalar field
fluctuation gives $\rr = (H/\dot{\phi}_0)(\delta\phi)_{k=aH} \sim
H^2/\dot{\phi}_0$.  Now we can express $H^2$ and $\dot{\phi}_0$ in
terms of the inflaton potential and its derivative.  The background
equation of motion from Section 2.3 in the slow-roll limit (ignoring
the $\ddot{\phi}$ term) gives $\dot{\phi} \sim \sub{V}/3H$, and the
Friedmann equation relates $H$ and $V(\phi)$: $H^2 =
8\pi\rho/3m_{Pl}^2 \sim V(\phi)/m_{Pl}^2$.  Substituting $V$ and
$\sub{V}$ gives the familiar result for the perturbation spectrum
first found in this manner by Bardeen, Steinhart, and Turner (1983),
\be
\label{eq:rough}
k^{3/2}\rr_k \sim A_S(k) \sim  \frac{1}{m_{Pl}^3} 
\frac{V^{3/2}(\phi)}{\sub{V}} .
\ee

Since the scale factor increases so rapidly during inflation, all
astrophysical scales of interest correspond to a rather narrow range
of inflaton field values.  For flat potentials, $V(\phi)$ and
$\sub{V}(\phi)$ does not change much during inflation, so one expects
$A_S(k)$ to be roughly independent of $k$.  This is the reason for the
often repeated ``result'' that inflation leads to an approximate
Harrison-Zel'dovich spectrum of scalar density perturbations.

\subsection{The Three Step Program for Better Predictions }

The calculation of $A_S(k)$ in Eq.\ (\ref{eq:rough}) was sufficiently
accurate for a decade.  But even with present-day data, and especially
looking forward to the wealth of information expected in the near
future, such as the angular power spectrum of CMB fluctuations up to
multipole number of more than $10^3$, more accurate predictions are
required.

As the result of considerable effort, in the past few years some
progress has been made in improving the accuracy of the calculation of
the density perturbation spectrum.  Let me describe three basic steps
in the road toward better accuracy:
\begin{enumerate}
\item a better treatment of the background classical dynamics by use of
 Hamilton-Jacobi formalism,
\item a better formalism of quantum corrections by use of the variational
approach, and
\item the calculation of the spectra in terms of slow-roll parameters.
\end{enumerate}

I have already discussed the advantage of treating $H$ as fundamental,
and parameterizing its evolution by $\phi$ rather than time. In
principle, the Hamilton--Jacobi formalism enables one to treat the
dynamical evolution of the scalar field exactly, at least at the
classical level. In practice, however, the separated Hamilton-Jacobi
equation, the first line of Eq.~(\ref{eq:hj}), is rather difficult to
solve. On the other hand, the analysis can proceed straightforwardly
once the functional form of the expansion parameter $H(\phi)$ has been
determined. This suggests that one should view $H(\phi)$ as the
fundamental quantity in the analysis (Lidsey 1991b, 1993).  This is in
contrast to the more traditional approaches to inflationary cosmology,
whereby the particle physics sector of the model --- as defined by the
specific form of the inflaton potential $V(\phi)$ --- is regarded as
the input parameter.

It proves convenient to express the scalar and tensor perturbation
spectra in terms of $H(\phi)$ and its derivatives.  The slow-roll
approximation is an expansion in terms of quantities derived from
appropriate derivatives of the Hubble expansion parameter. Since {\em
at a given point} each derivative is independent, there are in general
an infinite number of these terms, but only the first few enter into
any expressions of interest. The first three are defined as
\begin{equation}
\label{epsilon}
\epsilon (\phi) \equiv \frac{3\dot{\phi}^2}{2} \left[ V+\frac{1}{2} 
	\dot{\phi}^2 \right]^{-1} =\frac{m_{{\rm Pl}}^2}{4\pi} \left( 
	\frac{\sub{H} (\phi) }{H(\phi)} \right)^2 \,,
\end{equation}
\begin{equation}
\label{eta}
\eta (\phi)
\equiv -\frac{\ddot{\phi}}{H\dot{\phi}} = \frac{m_{{\rm Pl}}^2}{4\pi} 
	\frac{\subsub{H}(\phi)}{H(\phi)} = \epsilon -\frac{m_{{\rm Pl}} \, 
	\sub{\epsilon}}{\sqrt{16\pi \epsilon}} \,,
\end{equation}
\begin{equation}
\label{xi}
\xi (\phi) \equiv \frac{m_{{\rm Pl}}^2}{4\pi} \left( \frac{\sub{H}(\phi) 
\subsubsub{H} 
	(\phi)}{H^2(\phi)} \right)^{1/2} = \left[ \epsilon \eta -
	\left( \frac{m_{{\rm Pl}}^2 \, \epsilon}{4\pi} \right)^{1/2} 
	\sub{\eta} \right]^{1/2} \,.
\end{equation}
One need not be concerned as to the sign of the square root in the
definition of $\xi$; it turns out that only $\xi^2$, and not $\xi$
itself, will appear in our formulae. We emphasize
that the choice $\dot{\phi} > 0$ implies that $\sqrt{\epsilon} = -
\sqrt{m_{{\rm Pl}}^2/4\pi} \, \sub{H}/H$.

One can show that inflation ends when $\epsilon=1$.  The slow-roll
approximation, as I use it here, involves assuming $\{\epsilon,\
\eta,\ \xi\} $ are all less than unity.  This is somewhat more
restrictive than just saying that $H$ changes slowly enough for
inflation to occur: that only requires $\epsilon<1$.

Probably the most important advance is the development of the Mukhanov
formalism for the perturbation calculation.  Recall that the action
for the Einstein--scalar field system is 
\be S=-\int d^4\! x \sqrt{-g}
\left[ \frac{m_{Pl}^2}{16\pi}R - \frac{1}{2}(\nabla\phi)^2 +V(\phi)
\right], 
\ee 
with $g_{\mu\nu}= g_{\mu\nu}^{FRW} +\delta
g_{\mu\nu}(A,B,\psi,E)$ and $\phi = \phi_0(\tau) +\delta\phi({\bf
x},\tau)$.

Before quantizing the system, one must express the theory in terms of
the ADM variables, expand to second order in the perturbations, apply
the background field equations, and integrate by parts when judicious.
Now the procedure is quite long and tedious.  Details can be found in
the review article of Mukhanov, Feldman, and Brandenberger (1992).
When the dust settles, the variation of the action can be expressed in
terms of the dynamical variable $u=a(\delta\phi+\phi_0'\psi/{\cal
H})=z{\cal R}$, where $z=a\phi_o'/{\cal H} = a\dot{\phi}_0/H$: \be
\label{eq:sqft}
\delta_2S = \frac{1}{2}\int d^4\!x \left ( u'^2 -\delta^{ij}u,_iu,_j +
\frac{z''}{z}u^2 \right).  
\ee 
Now this is really remarkable, because
the complicated dynamics of scalar field perturbations coupled to
metric perturbations can be cast into the dynamics of a system we know
well: a scalar field $u$ in flat spacetime with (time-dependent and
negative) mass $m^2=-z''/z$.

Now scalar field theory in flat space is well understood.  So we can
use the tool of scalar quantum field theory as a sort of hammer to
pound out the answer.  Of course, we have to make sure we have the
right tool, because as the saying goes, ``when you have a hammer in
your hand, everything you see looks like a nail.''

\subsection{Quantization}

The quantization of the action in Eq.\ (\ref{eq:sqft}) is really
 rather straightforward: From the scalar field $u({\bf x},\tau)$, form
 the conjugate momentum $\pi({\bf x},\tau)$, and form the Hamiltonian
 from $u({\bf x},\tau)$ and $\pi({\bf x},\tau)$.  Then promote the
 classical field and its conjugate momentum to operators, $u({\bf
 x},\tau) \rightarrow \widehat{u}({\bf x},\tau)$, and $\pi({\bf
 x},\tau) \rightarrow \widehat{\pi}({\bf x},\tau)$ and impose the
 canonical equal-time commutation relations $[\widehat{u}({\bf
 x},\tau), \widehat{u}({\bf y},\tau)] = [\widehat{\pi}({\bf x},\tau),
 \widehat{\pi}({\bf y},\tau)] = 0$, and $[\widehat{u}({\bf x},\tau),
 \widehat{\pi}({\bf y},\tau)] = i \delta^3({\bf x} - {\bf y})$.

If one expands the field operator in Fourier modes associated with
creation and annihilation operators $\widehat{a}_k^\dagger$ and
$\widehat{a}_k$, then the field becomes 
\be \widehat{u}({\bf x},\tau)
= (2\pi)^{-3/2} \int d^3\!k \left[ u_k(\tau) \widehat{a}_k e^{ i{\bf
k} \cdot {\bf x} } + u_k^*(\tau) \widehat{a}^\dagger_k e^{-i{\bf k}
\cdot {\bf x}} \right].  
\ee 
The field equation for $u_k$ is the
familiar Klein--Gordan equation \be
\label{eq:kg}
u_k'' + \left( k^2 - \frac{z''}{z} \right) u_k = 0 .  
\ee 
Of course we
must specify the boundary condition.  In our case, we want $u_{k\gg
aH} \rightarrow e^{-ik\tau}/k^{1/2}$, i.e., plane-wave solutions.

It is pleasing to note that any solution to Eq.\ (\ref{eq:kg}) will
have the feature that well beyond the Hubble radius, ${\cal R}_k$ will
be constant.  Note that in the limit $k\rightarrow 0$ the field
equation becomes $u_k'' - (z''/z)u_k=0$, which obviously has solution
$u_k\propto z$.  Now since ${\cal R}_k = u_k/z$, on scales much larger
than the Hubble radius ${\cal R}_k \rightarrow$ constant.

\section{Tensor Perturbations}

\def\hij{{h_{ij}}}

In addition to the scalar perturbations in Eq.\ (\ref{eq:SCALARP}),
the most general metric contains perturbations that transform like a
tensor on the spatial hypersurfaces.  These tensor perturbations enter
the metric as \be
\label{eq:TENSORP}
ds^2= a^2(\tau)\left[ d\tau^2 - (\delta_{ij} + \hij) dx^idx^j\right] .
\ee 
As can be seen by explicit calculation from the Einstein
equations, the metric perturbation $\hij$ does not couple to the
stress-energy tensor, but describes the propagation of gravitational
waves.  The gravitational waves are not important for large-scale
structure, but they do have an effect of the CMB, at least for small
multipole number.

Since by construction $\hij$ is a transverse, traceless tensor, it has
two degrees of freedom, usually denoted as $h_\times$ and $h_+$.
(From the quantum view, gravitational waves are the propagating part
of the gravitational degrees of freedom, corresponding to a massless
spin-two particle, which of course has two degrees of freedom.)

Now just as was done for the scalar degrees of freedom, one
substitutes the metric Eq.\ (\ref{eq:TENSORP}) into the
Einstein--Hilbert action, and expands to quadratic order in $\hij$,
with the result
\be \delta_2S = \frac{m^2_{Pl}}{64\pi} \int d\tau
d^3\! x \, a^2(\tau) \, \partial_\mu h^i_{~j} \partial^\mu h_i^{~j}.
\ee 
Since our goal is quantization, and we know how to quantize scalar
field theory, we want to make $\delta_2S $ look as much as possible
like the action for a scalar field.  To this end, it is very
convenient to define the rescaled variable $P^i_{~j}(x) =
(m^2_{Pl}/32\pi)^{1/2} a(\tau) h^i_{~j}(x)$.  In terms of
$P^i_{~j}(x)$, $\delta_2S $ becomes 
\be \delta_2S = \frac{1}{2}\int
d\tau \, d^3\!x \, \left[ \left(\partial_\tau P_i^{~j} \right)
\left(\partial_\tau P^i_{~j} \right) - \delta^{mn} \left(\partial_m
P_i^{~j} \right) \left(\partial_n P^i_{~j} \right) + \frac{a''}{a}
P_i^{~j} P^i_{~j} \right] .  
\ee 
This may be interpreted as the action
for two scalar fields in Minkowski spacetime, each with effective mass
squared $-a_{\tau\tau}/a$.  We can now proceed with quantization
exactly as in the scalar case.

Again, just as in the scalar case for ${\cal R}$, we perform a Fourier
decomposition of $P^i_{~j}$.  But since there are two degrees of
freedom, we must include a polarization tensor ${\epsilon^i}_j ({\bf
k}; \lambda )$, which satisfies the conditions $\epsilon_{ij}
=\epsilon_{ji}$, $\epsilon_{ii}=0$, $k^i\epsilon_{ij}=0$ and
${\epsilon^i}_j ({\bf k},\lambda ) {\epsilon_i}^{j*} ({\bf k},
\lambda' ) =\delta_{\lambda\lambda'}$. The analysis is further
simplified if we choose $\epsilon_{ij}(-{\bf k}, \lambda
)=\epsilon_{ij}^* ({\bf k} ,\lambda )$.  The Fourier decomposition can
be written as 
\be P^i_{~j} = \sum_{\lambda=\times, +} \int \frac{d^3\!
k}{(2\pi)^{3/2}} \ \ v_{k,\lambda} \ \epsilon^i_{~j}(k;\lambda)\
e^{i{\bf k} \cdot {\bf x}} .  
\ee 
In terms of $v_k$, the spectrum of
gravitational waves, $A_T(k)$, is defined as
\begin{equation}
\frac{k^3}{2\pi^2} \langle {v}_{{\bf k},\lambda } {v}^*_{{\bf l},\lambda} 
\rangle =\frac{10^2m_{{\rm Pl}}^2 a^2}{32\pi} \,
A_T^2(k) \, \delta^{(3)} ({\bf k}-{\bf l}) \,.
\end{equation}

Now returning to the quantization of the perturbations,  in momentum 
space the tensor perturbation action is 
\begin{equation}
\label{rewrite}
\delta_2S = \frac{1}{2} \sum_{\lambda =+,\times} \int d\tau d^3\! k
\left[ \left( \partial_{\tau} \left| v_{k,\lambda} \right| \right)^2 
-\left( k^2 -\frac{a''}{a} \right) \left| v_{ k,\lambda} 
\right|^2 \right] \,.
\end{equation}
We can now quantize $v_{k,\lambda}$ in the usual way, promoting the field 
to an operator with canonical quantization conditions.

The mode equation for $v_k$ becomes
\begin{equation}
\label{fieldeqn}
\frac{d^2v_k}{d\tau^2} +\left( k^2 -\frac{a''}{a} \right) 
v_k =0 \,.
\end{equation}
This equation can be compared to the mode equation for scalar
perturbations, Eq.\ (\ref{eq:kg}).  The mode equation is somewhat
simpler than the scalar case because $a''/a =2a^2H^2 ( 1 -
\epsilon/2)$ is generally a simpler function than $z''/z$.

\section{A Variety of Models  (Some Realistic, Others Illustrative)}

\subsection{Solution Procedure}
The procedure is simple (in principle): solve Eq.\ (\ref{eq:kg}) for
$u_k$, then find ${\cal R}_k = u_k/z$ to give $A_S(k)$, which together
with a transfer function, yields the power spectrum which can be
compared to observations.  Then solve Eq.\ (\ref{fieldeqn}) for $v_k$,
to give $A_T(k)$.

The trouble is that exact solutions to the wave equations are hard to find, 
partly because the mass terms are so complicated:
\begin{eqnarray}
\label{zderiv}
\frac{z''}{z} & = & 2 a^2H^2\left[  1 + \epsilon -\frac{3}{2} \eta + 
\epsilon^2 - 2\epsilon\eta +\frac{1}{2}\eta^2 + \frac{1}{2}\xi^2  \right],  
\cr
\frac{a''}{a} & = & 2a^2H^2\left( 1- \frac{1}{2}\epsilon \right)
\end{eqnarray}
where the $\tau$ dependence of $H,\ \epsilon,\ \eta,$ and $\xi$ are
found from their dependence upon $\phi$.  In fact, only two exact
solutions of Eqs.\ (\ref{eq:kg}) and (\ref{fieldeqn}) are known.  The
first is a power-law solution found by Stewart and Lyth (1993), and
the second, yet unnamed, has been found by Easther (1996).

The first step is to express the conformal time, $d\tau=dt/a(t)$ in
terms of $aH$ and the slow-roll parameters.  In general the result is
\be 
\tau = \int \frac{da}{a^2H} = -\frac{1}{aH} + \int
\frac{\epsilon}{a^2H}\ da.  
\ee 
{\it If $\epsilon$ is constant,} then
$\tau^{-1} = -aH(1-\epsilon)$ ($\tau$ is negative during inflation,
with $\tau=0$ corresponding to the infinite future). If $\epsilon$ is
not constant, then integrating by parts an infinite number of times,
one can obtain
\begin{equation}
\label{eq:tau}
\tau =-\frac{1}{aH}\frac{1}{1-\epsilon} - \frac{2\epsilon \zeta}{aH} + 
\mbox{expansion in slow-roll parameters $\zeta$ etc.} \,,
\end{equation}
where $\zeta = \epsilon -
\eta$, and $\epsilon$ can now have arbitrary time dependence. 

In the next section I will review the exact power-law solution, and
and the section after that I will discuss how to use that exact
solution to construct perturbative solutions for other models.

\subsection{Power-Law Inflation}

In the power-law model the Hubble parameter is expressed in terms of
the Planck mass and a parameter $p$: $H(\phi) \propto \exp
\sqrt{4\pi\phi^2/pm_{Pl}^2} $, which results from a scalar potential
of the form $V(\phi)\propto e^\phi$ (Abbott and Wise 1984, Lucchin
and Matarrese 1985).  Obviously this type of potential
is not a fundamental, renormalizable scalar potential, but it is the
type of effective low-energy potential for dilaton-like degrees of
freedom in string theories and Kaluza-Klein theories.

For $H(\phi)\propto e^\phi$, $\epsilon$, $\eta$, and $\xi$ will be
equal and {\it constant:} $\epsilon = \eta = \xi = p^{-1}$.

Now one can proceed to find $z''/z$ and $a''/a$, with the result
$z''/z=(\nu^2-1/4)/\tau^2$ and $a''/a= (\mu^2-1/4)/\tau^2$, where $\nu
= (3/2) + (p-1)^{-1}$ and $\mu = (3/2) + (p-1)^{-1}$ (For power-law
inflation $\nu$ and $\mu$ coincide, though in general they do not.)

For power-law inflation the mode equations are simply a Bessel equation:
\begin{eqnarray}
\label{bessel}
\left [  \frac{d^2}{d\tau^2} + k^2 - \frac{\nu^2-4^{-1}}{\tau^2}\right] u_k 
&= &0  \cr
\left [  \frac{d^2}{d\tau^2} + k^2 - \frac{\mu^2-4^{-1}}{\tau^2}\right] v_k 
&= &0,
\end{eqnarray}
which for the boundary conditions we impose are solved by
$H_\nu^{(1)}(-k\tau)$ and $H_\mu^{(1)}(-k\tau)$, Hankel functions of
the first kind of order $\nu$ and $\mu$.

We are interested in the asymptotic forms of $u_k/z$ and $v_k$ for $k\ll aH$,
 which are easily found to be
\begin{eqnarray}
\label{modelimit}
u_k & \rightarrow & e^{i(\nu -1/2)\pi /2} 2^{\nu -3/2} 
\frac{ \Gamma (\nu)}{\Gamma (3/2)} \frac{1}{\sqrt{2k}} (-k \tau )^{-\nu 
+1/2}  \cr
v_k & \rightarrow & {\mbox {above with}}\  \nu \rightarrow \mu
\end{eqnarray}
which yields $A_S(k) \propto H^2/|H'| $ and $A_T(k) \propto H$, with
both expressions evaluated at $k=aH$.  Now using the fact that at
Hubble radius crossing $H(\phi) \propto k^{1/p}$ from Eq.\
(\ref{eq:kphi}), we find a power-law spectrum $A_S(k)$ and $A_T (k)$
proportional to $ k^{-1/p} $.

The scalar spectral index is defined as $n(k) - 1 = d\ln A_S^2/d\ln
k$.  Writing $A_S^2(k) \propto k^{-2/p} $ the above power-law spectrum
gives $n-1 = -2/p$, a departure from the $n=1$ Harrison-Zel'dovich
result.  Defining the tensor spectral index, $n_T(k)$ as $n_T(k) =
d\ln A_T^2/d\ln k$, for power-law inflation $n_T = -2/p$.

\subsection{General Potentials}

After working hard to find an exact solution, we can now make an
expansion about it for general potentials.  The power-law inflation
case corresponded to the slow-roll parameters being equal, and hence
exactly constant.  In general they can be different, which means they
will pick up a time dependence.

Assuming that $\epsilon$, as well as $\zeta=\epsilon-\eta$ are small, then 
Eq.\ (\ref{eq:tau}) can be approximated to give $\tau = -(1+\epsilon)/(aH)$.

Having this expression for $\tau$, we can now immediately use
Eq.~(\ref{zderiv}), which must also be truncated to first-order. This
gives the same Bessel equation Eq.~(\ref{bessel}), but now with $\nu$
given by $\nu = 3/2 + 2\epsilon - \eta $ and $\mu$ given by $\mu = 3/2
+ \epsilon $ .  The assumption that treats $\epsilon$ as constant also
allows $\eta$ to be taken as constant, but crucially, $\epsilon$ and
$\eta$ need no longer be the same since we are consistent to
first-order in their difference. The differences between further
slow-roll parameters and $\epsilon$ lead to higher order effects, and
so incorporating $\epsilon$ and $\eta$ in this manner is applicable to
an arbitrary inflaton potential to next-order.  The same solution
Eq.~(\ref{modelimit}) can be used with the new form of $\nu$, but for
consistency it should be expanded to the same order. This gives the
final answer, which is true for general inflation potentials to this
order, of (Stewart \& Lyth 1993)
\begin{eqnarray}
\label{2ndscal}
A_S(k) & = &  \frac{2}{5\sqrt{\pi}}\  \left\{ \left[ 1- (2C+1) 
\epsilon(\phi) + C \eta(\phi)
	\right] \;  \frac{1}{\sqrt{\epsilon(\phi)}} 
	\frac{H(\phi)}{m_{Pl}} \right\}_{k=aH} \  , \cr
A_T(k) & = &  \frac{2}{5\sqrt{\pi}}\  
\left\{ \left[  1 - (C+1)\epsilon(\phi) \right] \frac{H(\phi)}{m_{Pl}} 
\right\}_{k=aH} \ \ .
\end{eqnarray}
where $C = -2 +\ln 2 +\gamma \simeq -0.73$ is a numerical constant,
$\gamma$ being the Euler constant originating in the expansion of the
Gamma function.   Of particular interest is the ratio
\be
\frac{A_T^2(k)}{A_S^2(k)} = \epsilon(\phi) \  [1+2C(\epsilon(\phi) - 
\eta(\phi)) ] \ \ .
\ee

It is useful once again to point out that the $\phi
\longleftrightarrow k$ connection is made through Eq. (\ref{eq:kphi}),
which can be written in the form 
\be \frac{d\ln k}{d\phi} =
\sqrt{\frac{4\pi}{m_{Pl}^2}}\ \ \frac{\epsilon-1}{\sqrt{\epsilon}}.
\ee

For the spectral indices $n(k)-1\equiv d \ln A_S^2(k)/d \ln k$ and
$n_T(k)\equiv d \ln A_T^2(k)/d \ln k$, it is easy to show that
\begin{eqnarray}
n(k) & = & 1 - 4\epsilon  + 2\eta - [8(C+1)\epsilon^2 - (6+10C)\epsilon\eta 
+2C\xi^2 ], \cr
n_T(k) & = & -2\epsilon[ 1 + (3+2C)\epsilon - 2(1+C)\eta].
\end{eqnarray}
Obviously, the usual Harrison-Zel'dovich result $n=1$ is obtained if
the slow-roll parameters $\{\epsilon,\, \eta,\, \xi\}$ are all much
less than unity.  But recall that $\epsilon=1$ defines the end of
inflation, so there is no reason to assume that the slow-roll
parameters must be much less than unity 50 e-folds from the end of
inflation.

\begin{table}
\begin{center}
\begin{tabular}{c|c|c}
\hline \hline
observable & lowest-order & next-order \\
\hline \hline
 & & \\
$A_T^2(k_0)$		& $H(\phi_0)$	
					& $H(\phi_0)$, $\epsilon(\phi_0)$ \\
$A_S^2(k_0)$		& $H(\phi_0)$, $\epsilon(\phi_0)$	& 
$H(\phi_0)$, $\epsilon(\phi_0)$, $\eta(\phi_0)$ \\
$A_T^2(k_0)/A_S^2(k_0)$ & $\epsilon(\phi_0)$& $\epsilon(\phi_0)$, 
$\eta(\phi_0)$\\
$n_T(k_0)$     & $\epsilon(\phi_0)$& $\epsilon(\phi_0)$, $\eta(\phi_0)$\\
$1-n(k_0)$		& $\epsilon(\phi_0)$, $\eta(\phi_0)$	& 
$\epsilon(\phi_0)$, $\eta(\phi_0)$, $\xi(\phi_0)$\\
\end{tabular}
\end{center}
Table 1: The observables, $A_T^2$, $A_S^2$, $n$, and $n_T$ at the
point $k_0$ may be expressed in terms of $H$ and the slow-roll
parameters at the point $\phi_0$.  This table lists the inflation
parameters required to predict the observable to the indicated order.
\end{table}

\subsection{The Consistency Relation}
Before turning to specific models, it is important to recognize a
``consistency'' relation.  The overall amplitude is a free parameter
determined by the normalization of the expansion rate $H$ during
inflation (or equivalently the scalar field potential $V$). On the
other hand, the relative amplitude of the two spectra is given to
lowest order by
\begin{equation}
\label{rat1}
\frac{A_T^2}{A_S^2} = \epsilon \,.
\end{equation}
Thus, to lowest order in the slow-roll parameters, there exists a
simple relationship between the relative amplitude and the tensor
spectral index:
\begin{equation}
\label{firstconsistency}
n_T = - 2 \frac{A_T^2}{A_S^2} \,.
\end{equation}
This is the lowest-order consistency equation and represents an
extremely distinctive signature of inflationary models. It is
difficult to conceive of such a relation occurring via any other
mechanism for the generation of the spectra.

Since it is possible for the spectra to have different indices, the
assumption that their ratio is fixed can be true only for a limited
range of scales, but the correction enters at a higher order in the
slow-roll parameters.

\subsection{Other Models}
Here I briefly give some results {\it to lowest order} in the
slow-roll parameters for the spectral index in a couple of
well-studied inflation models.  I will work out polynomial chaotic
inflation in detail, and only describe the other models and give the
results.

Now in this section we are treating the potential as input, so it is
useful to have the lowest-order results for the slow-roll parameters
in terms of $V$.  These were studied by Kolb and Vadas (1994), with
the result 
\be \epsilon = \frac{m_{Pl}^2}{16\pi}\left(
\frac{\sub{V}}{V}\right)^2 \quad {\rm and} \quad \eta =
-\frac{m_{Pl}^2}{16\pi}\left( \frac{\sub{V}}{V}\right)^2 +
\frac{m_{Pl}^2}{8\pi}\left( \frac{\subsub{V}}{V}\right) .  
\ee 
I will use the lowest-order result $n=1 - 4\epsilon + 2\eta$ and
$A_T^2/A_S^2=-n_T/2=\epsilon$.

Now of course the slow-roll parameters are a function of $\phi$, which
implies they are a function of $k$.  But since we are working to
lowest order, we can assume that the spectral indices are constant,
and the values associated with Hubble radius crossing about 50 e-folds
from the end of inflation.  Generally we will have to find the value
of the field 50 e-folds from the end of inflation.  We will denote
this as $\phi_{50}$.

The end of inflation is defined by $\epsilon(\phi)=1$, and the
definition of the number of e-folds from the end of inflation is \be
N(\phi) = \int_a^{a_{end}} \frac{da}{a} = \int_t^{a_{end}} H\ dt =
\int_\phi^{\phi_{end}} \frac{H}{\dot{\phi}} d\phi =
\sqrt{\frac{4\pi}{m_{Pl}^2}} \int_\phi^{\phi_{end}}
\frac{d\phi}{\sqrt{\epsilon(\phi)}}\ \ .  
\ee

\subsubsection{Polynomial Chaotic Inflation}
Probably because of its simplicity, the most popular inflation model
is polynomial chaotic inflation, where the potential is assumed to be
$V(\phi)=\phi^p$.  A potential of this form has been championed by
Linde.

With the potential in this form, the first two slow-roll parameters
are 
\be \epsilon = \frac{m_{Pl}^2}{16\pi} \ \frac{p^2}{\phi^2} \quad
{\rm and} \quad \eta = - \frac{m_{Pl}^2}{16\pi} \ \frac{p^2}{\phi^2} +
\frac{m_{Pl}^2}{8\pi}\ \frac{p(p-1)}{\phi^2}.  
\ee

The end of inflation is found by setting $\epsilon=1$, which gives
$\phi_{end}^2/m_{Pl}^2 = p^2/16\pi$.  For this model $N(\phi) = 4\pi
(\phi^2 - \phi_{end}^2)/pm_{Pl}^2$, which gives $\epsilon =
pm_{Pl}^2/16\pi\phi_{50}^2 = p/(p+200)$, and $\eta = (p-2)/(p+200)$.

Finally, the above values of $\epsilon$ and $\eta$ give 
\begin{eqnarray}
n & = &1 - 4\epsilon + 2\eta = 1 - \frac{4 + 2p}{p + 200} \sim 1 -
\frac{p+2}{100} \cr A_T^2/A_S^2 & = &\epsilon = \frac{p}{p + 200}
\end{eqnarray}

This gives a flavor of the calculations that can easily be done for
the other inflation models discussed in the following subsections.
The results are given in Table 1.

\subsubsection{Power Law Inflation}  

We have already discussed the power-law inflation model.  In that
model $\epsilon=\eta=1/p$.  Of course the fact that $\epsilon$ is a
constant means that some other machinery must be introduced for the
highly desirable result of an end to inflation.

\begin{table}
\begin{center}
\begin{tabular}{c|c|c}
\hline \hline
model & $n$ & $A_T^2/A_S^2$  \\
\hline \hline & & \\ 
power-law:& &\\
$ \exp \sqrt{16\pi\phi^2/pm_{Pl}^2}$   & $1-2/p$   & $1/p$    \\
$p=5$                  & 0.6     & 0.2    \\
$p=10$                  & 0.8   & 0.1    \\ 
\hline
polynomial chaotic: & \\
$\phi^p$  & $1-(p+2)/100$	&$p/(p+200)$ \\  
$p=2$ 	& 0.96			& 0.01 \\
$p=4$ 	& 0.94 			& 0.02 \\
\hline
natural  & \\
$ 1\pm \cos (\phi/f) $ & $ 1-m_{Pl}^2/8\pi f^2$ & 
$(m_{Pl}^2/16\pi f^2) \exp(-15m_{Pl}^2/2\pi f^2) $\\
$f^2=5m_{Pl}^2/8\pi $	& 0.8   & $6\times10^{-7}$  \\
$f^2=5m_{Pl}^2/16\pi $	& 0.6   & $8 \times10^{-12}$  \\
\hline
$R^2$ &\\   $ \{1-\exp[-(16\pi/3m_{Pl}^2)^{1/2}\phi]\}^2$ & 0.96 & 
$10^{-3}$  \\ 
\hline
CDM ($V=$ ????) & 1 & 0                            
\end{tabular}
\end{center}
Table 2: Well studied inflation models to lowest order.  The result
$A_T^2/A_S^2=(1-n)/2$, true for power-law inflation, is often
(incorrectly) used as a general result.  The relative contribution to
tensor modes to the CMB power spectrum for small multipole number is
approximately $6.5A_T^2/A_S^2$.
\end{table}

\subsubsection{Natural Inflation}
Natural inflation is a local Fermilab favorite (Freese, Frieman, and
Olinto, 1990).  In this model the potential takes the form of the
potential for a pseudo-Nambu-Goldstone boson: 
\be V(\phi) = \Lambda^4\left[ 1\pm cos(\phi/f)\right], 
\ee 
where $\Lambda$ and $f$
are mass scales.  The mass scale $f$ corresponds to the scale of the
breaking of the original $U(1)$ symmetry, and $\Lambda$ is the mass
scale associated with an explicit breaking term.  It is attractive to
consider $f$ to be of order $m_{Pl}$ and $\Lambda$ of order the GUT
scale.

Natural inflation is a great example of a model with a
non-renormalizable scalar potential.  Even though the underlying
theory may be renormalizable, there is no reason to expect that the
effective low-energy inflaton potential should be restricted to be of
a renormalizable form.

\subsubsection{$R^2$ Inflation}
$R^2$ inflation is actually the first model for inflation (Starobinsky
1980).  In this model the inflaton potential is not a fundamental
scalar field, bt has an origin in the gravity sector.  If one adds a
term quadratic in the Ricci scalar to the Einstein--Hilbert action,
\be
S = -\frac{1}{16\pi G} \int d^4\!x\ \sqrt{\widehat {g}}
\left [ R + \frac{R^2}{6M^2}  \right],
\ee
then by means of a conformal redefinition of the metric the $R^2$ term
can be eliminated.  Or, more precisely, the extra degree of freedom
can be rewritten to look like a minimally coupled scalar field with
action
\be
S = -\frac{1}{16\pi G} \int  d^4\!x\  \sqrt{\widehat{g}}\ \widehat{R} 
+ \int d^4\!x\  \sqrt{\widehat{g}}\ \left [ \frac{1}{2} \widehat{g}^{\mu\nu} 
\partial_\mu\phi 
\partial_\nu\phi - V(\phi) \right],
\ee
where 
\be
V(\phi) = \frac{3m^2_{Pl}M^2}{32\pi} \left\{ 1 - 
\exp \left[  -(16\pi/3m^2_{Pl} )^{1/2}\phi\right]   \right\}^2  .
\ee

Here is an example of an effective inflaton potential where the scalar
field need not be regarded as a fundamental scalar field degree of
freedom.  This suggests that the scalar field analysis described in
this paper may be useful for a class of models larger than just scalar
field models.

\section{So What's Your Point?}
In this lecture I have tried to make several points:
\begin{description}

\item 1. In one-field, slow-roll models of inflation it is possible to
make sufficiently accurate predictions of the observable parameters
such as $A_S$, $A_T$, $N$, and $n_T$.

\item 2. The restriction of ``one-field, slow-roll'' may not be as
restrictive as first imagined, because many models of inflation can be
written in this way even if they do not involve a fundamental scalar
field to start with.

\item 3.  Different models make different predictions for the
observables.  Soon it will be possible to sort through the models and
start weeding out those not in agreement with observation.

\item 4.  There is a consistency relation for these models, although
it may be difficult to check observationally.

\item 5.  Although not discussed in this lecture, with a little work
one can rework the expressions for the observables to express the
potential in terms of the observables.  Therefore, one might be able
to glean some information about a scalar field potential at energy
scales of about $10^{16}$GeV from astronomical observations (Copeland
et al.\ 1993a, 1993b, 1994; Lidsey et al.\ 1997; Turner 1993a, 1993b).

\end{description}

\section*{Acknowledgements}
My work is supported at Fermilab by the DOE and NASA under Grant NAG
5--2788. I am grateful for many helpful discussions with Scott
Dodelson, David Lyth, Michael Turner, and Will Kinney.  I am also most
grateful to my collaborators on reconstruction: Jim Lidsey, Andrew
Liddle, Ed Copeland, Tiago Barreiro, and Mark Abney.

\section*{References}

\begin{description}
\item Abbott, L. F., and M. Wise, 1984, Nucl. Phys. B {\bf 244}, 541.

\item Bardeen, J. M., 1980, Phys. Rev. D{\bf 22}, 1882.

\item Bardeen, J. M., P. J. Steinhardt and M. S. Turner, 1983, 
	Phys. Rev. D{\bf 28}, 679.

\item Copeland, E. J., E. W. Kolb, A. R. Liddle and J. E. Lidsey, 1993a, 
	Phys. Rev. Lett. {\bf 71}, 219.

\item Copeland, E. J., E. W. Kolb, A. R. Liddle and J. E. Lidsey, 1993b, 
	Phys. Rev. D{\bf 48}, 2529.

\item Copeland, E. J., E. W. Kolb, A. R. Liddle and J. E. Lidsey, 1994, 
	Phys. Rev. D{\bf 49}, 1840.

\item Easther, R., 1996, Class. Quant. Grav. {\bf 13}, 1775.

\item Freese, K., J. A. Frieman, and A. V. Olinto, 1990, Phys. Rev. Lett. 
{\bf 65}, 3233.

\item Grishchuk, L. P. and Yu. V. Sidorav, 1988, in Fourth Seminar on 
	Quantum Gravity, eds M. A. Markov, V. A. Berezin and V. P. 
	Frolov (World Scientific, Singapore).

\item Kolb, E. W. and M. S. Turner, 1990, {\em The Early Universe}, 
	(Addison-Wesley, Redwood City, California).

\item Kolb, E. W. and Vadas, S. L.,  1994 Phys. Rev. D{\bf 50}, 2479.

\item Lidsey, J. E., 1991a, Phys. Lett. {\bf B273}, 42.

\item Lidsey, J. E., 1991b, Class. Quant. Grav. {\bf 8}, 923.

\item Lidsey, J. E., 1993, Gen. Rel. Grav. {\bf 25}, 399.

\item LidseyJ. E., A. R. Liddle, E. W. Kolb, E. J. Copeland, T. Barreiro, 
and M. Abney (L$^2$KCBA), 1997,  Rev. Mod. Phys. (to appear, April 1997).

\item Liddle, A. R. and D. H. Lyth, 1993, Phys. Rept. {\bf 231}, 1.

\item Lucchin, F. and S. Matarrase, 1985, Phys. Rev. D{\bf 32}, 1316.

\item Mukhanov, V. F., H. A. Feldman and R. H. Brandenberger, 1992, 
	Phys. Rept. {\bf 215}, 203.

\item Muslimov, A. G., 1990, Class. Quant. Grav. {\bf 7}, 231.

\item Salopek, D. S. and J. R. Bond, 1990, Phys. Rev. D{\bf 42}, 3936.

\item Salopek, D. S. and J. R. Bond, 1991, Phys. Rev. D{\bf 43}, 1005.

\item Starobinsky, A. A., 1980, Phys. Lett. {\bf B91}, 99.

\item Stewart, J. M., 1990, Class. Quant. Grav. {\bf 7}, 1169.

\item Stewart, E. D. and D. H. Lyth, 1993, Phys. Lett. {\bf B302}, 171.

\item Turner, M. S., 1993a, Phys. Rev. {\bf D48}, 3502.

\item Turner, M. S., 1993b, Phys. Rev. {\bf D48}, 5539.

\end{description}

\end{document}